# Graphene Nano-Ribbon Electronics

Zhihong Chen[*], Yu-Ming Lin, Michael J. Rooks and Phaedon Avouris

IBM T.J. Watson Research Center, Yorktown Heights, NY 10598, USA

___

**Abstract**

We have fabricated graphene nano-ribbon field-effect transistor devices and investigated their electrical properties as a function of ribbon width. Our experiments show that the resistivity of a ribbon increases as its width decreases, indicating the impact of edge states. Analysis of temperature dependent measurements suggests a finite quantum confinement gap opening in narrow ribbons. The electrical current noise of the graphene ribbon devices at low frequency is found to be dominated by the *1/f* noise.

___

## 1. Introduction

Graphene, a single layer of graphite, has been used extensively as a basis for the discussion of the electronic structure of carbon nanotubes (CNTs) [1,2]. The latter can be considered as resulting from the folding of a graphene ribbon to form a seamless cylinder. However, the graphene itself is a 2D, zero-gap semiconductor with extremely interesting electronic properties. The linear, light-like, relationship of the electronic energy $E_k$ and the 2D momentum $k=(k_x, k_y)$, i.e. $E_k=\upsilon_F k$, implies that the electron effective mass is zero and the system can be described by a relativistic Dirac equation where the role of the velocity of light is played by the Fermi velocity $\upsilon_F$. It was not however, until relatively recently that single graphene layers were produced [3] and became the subject of intense study [4-7]. These studies revealed remarkable transport properties including electron and hole mobilities of the order of $10^4$ cm$^2$/V.s, i.e. approaching those reported for single CNTs [8]. This has raised the possibility of using graphene in device applications in a manner similar to CNTs. Being, however, a zero-gap semiconductor, graphene cannot be used directly in applications such as field-effect transistors (FETs). However, in addition to the 2D confinement, the graphene electrons can be further confined by forming narrow ribbons, e.g quantizing $k_y$. The width confinement is expected to result in a split of the original two-dimensional (2D) energy dispersion of graphene into a number of one-dimensional (1D) modes. Depending on the boundary conditions, some sets of these 1D modes do not pass through the intersection point of the conduction and valence band, and these quasi-1D graphene ribbons become semiconductors with a finite energy gap. The properties of GNRs would be quite different from those of graphene, for example the carrier mobility is expected to decrease as the gap increases [9].

In this report we describe the field switching and transport at different temperatures in narrow GNRs produced by electron beam lithography and etching techniques. Ribbons as narrow as 20 nm have been measured. Like CNTs, GNRs have defects that can scatter the carriers. These defects can be structural, chemical, or charged substrate sites. Furthermore, unlike CNTs, where periodic boundary conditions are present, GNRs have edges with localized states [10] that can also affect transport. As very narrow GNRs are needed to achieve the gap of even large diameter CNTs, the effect of the edges can be critical. Another question that is connected to scattering is the issue of electrical noise. Noise is known to increase with decreasing size, as described by Hooge's rule [11]. In the case of CNTs it was found that the dominant form of noise is *1/f* noise and its origin was ascribed to charge fluctuations involving substrate traps [12]. It is therefore, important to find if the same holds true for GNRs.

## 2. Experimental

Graphene sheets were extracted by micromechanical cleavage [3] from three-dimensional highly ordered pyrolytic graphite (HOPG) and deposited onto heavily p-doped Si substrates covered with a *200 nm* SiO$_2$ layer. Atomic force microscopy (AFM) was used to measure the thickness of the sheet to identify whether it is a single or few layer graphene. The graphene was patterned with e-beam lithography, followed by an oxygen plasma etching process in which the e-beam resist, HSQ, was used as the etching mask, forming GNRs with various widths. As shown in the SEM image in Fig. 1, palladium (Pd) source drain contacts were deposited on top of the GNR, forming a three terminal field-effect transistor (FET) device with the Si substrate used as the back gate. All devices fabricated have a channel length of 1μm, and the width of the GNR studied ranges from *20 nm* to *500 nm*.

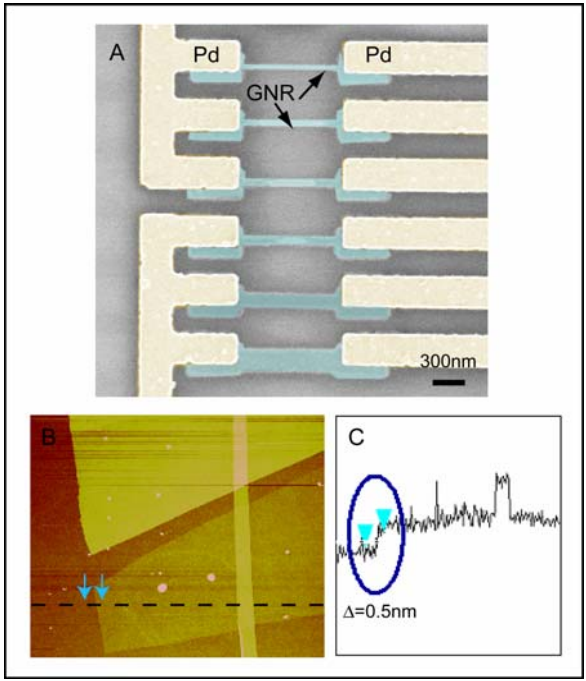

Fig. 1. (A) SEM picture of GNR devices fabricated on a *200 nm* SiO$_2$ substrate. The widths of the GNRs from top to bottom are *20 nm*, *30 nm*, *40 nm*, *50 nm*, *100 nm* and *200 nm*. (B) AFM image of a single layer graphene before lithographic process. (C) Cross-section measurement of the AFM, which provides the thickness of the graphene. When accounting the background noise and interaction between the graphene and substrate, we consider sheets thinner than *0.5 nm* to be single layer graphene.

### 3. Results and Discussion

All GNR devices were measured first at room temperature and showed current variation as a function of the Si back gate voltage. An example of current vs. back gate voltage for a *50 nm* ribbon is shown in the inset of Fig. 2. The "V" shaped current variation is observed in all of our GNR devices and has also been observed before in unconfined graphene devices [4,5]. It is expected that in a low carrier density system, the change of the carrier density, which is associated with the shift of the Fermi level, can be large enough to be readily observable. In graphene, the current reaches a minimum when the Fermi level is at the intersection of the conduction and valence band where the density-of-states is zero. In an undoped graphene layer, with a silicon back gate, the Fermi level is expected to be located in the vicinity of this intersection when no gate voltage is applied. In our experiments, we observe the conductivity minima to occur at different gate voltages for different devices. Even in the same device, the minimum can shift to different voltages when the device undergoes thermal cycling or gets exposed to different environments. Thus, although there may be an intrinsic conductivity minimum for an ideal graphene sheet [13-16], it is clear that our supported GNRs behave in a manner similar to CNT devices [8]; the observed hysteresis is due to trapped charges in the substrate (SiO$_2$). These charges act as electrostatic gating, and are sensitive to temperature and environment.

The combination of lithography and etching allows us to confine the width of the graphene and thus to fabricate the narrowest GNRs up to date. We have evaluated the resistivity of our GNRs at the minimum current point, i.e. the maximum resistivity, $\rho_{max}$. We note that these data were recorded at room temperature from freshly fabricated devices; in these first measurements trapped charges have not yet accumulated in the oxide. We found that for GNRs narrower than about *50 nm*, $\rho_{max}$ at room temperature increases as the width of the ribbon decreases. The increasing trend suggests that the impact from boundaries is no longer negligible when the size of the ribbon is sufficiently small. We consider two possible mechanisms which can be responsible for this trend. The first mechanism is the scattering occurring at rough boundaries. State-of-the-art lithographical and etching techniques do not allow atomic level resolution control and result in patterned lines with roughness on the order of a few nanometers. The narrower the ribbon is, the larger the contribution of the rough region to the total channel width. Therefore, more scattering induced resistivity shows up in the transport properties of narrow ribbons. Another explanation that assumes imperfection of the ribbon on an atomic scale would explain the resistivity increase as well. At the edges of the confined ribbons, the continuity of the hexagonal lattice is interrupted and the broken symmetry and change in bonding can also play an important role on the transport properties. It has been predicted sometime ago that an edge-state exists in finite graphene networks with a zigzag edge [10]. The flat band nature of the edge-state results in a peak in the local density of states near the Fermi level. Even for graphene systems with less developed zigzag edges, the theory predicts the survival of edge-states. Mixed edge states have been experimentally observed in scanning tunneling microscopy and spectroscopy even when the dangling bonds at the carbon edge have been saturated [17,18]. Considering the lack of perfect lithographic control on the lattice orientation as well as line roughness, it is very likely that localized, mixed edge states exist in our graphene ribbons and impact the electrical properties of the sample.

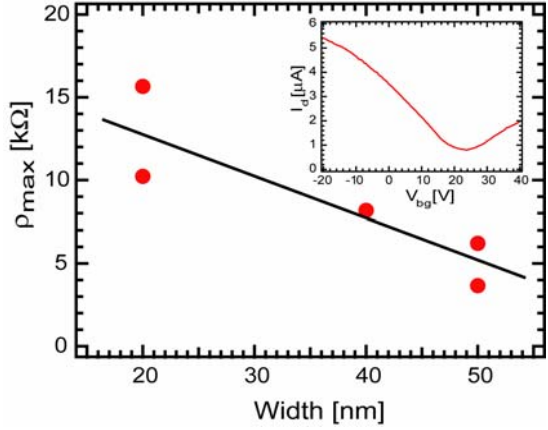

Fig. 2: Maximum resistivity of GNRs as a function of ribbon width. An example current vs. Si back gate curve for a *50 nm* GNR is shown in the inset. The measurement was taken at room temperature with a drain bias $V_{ds}=-0.1V$.

We also measured the electrical characteristics of the devices as a function of temperature. GNRs narrower than about *40 nm* show a distinct change of their electrical characteristics with temperature while wider ribbons do not. Fig. 3 shows the comparison between a *100 nm* and a *20 nm* ribbon. The minimum current of the *100 nm* GNR differs by less than a factor of 2 between the *300 K* and *4 K* data set. Different from this behavior, within the same gate voltage span, the *20 nm* GNR shows only about a factor of 2 current variation at *400 K*, but exhibits more than 1.5 orders of magnitude variation at *4 K*. This difference clearly states that the confinement in the *20 nm* ribbon opens a finite semiconducting gap in graphene. The gap is rather small and the off-state of the semiconductor is deteriorated by thermal carriers at high temperatures and only appears at low enough temperatures. The off-state current could have been further decreased if a smaller bias would have been used, that was however unavailable with the measurement tool we used (Agilent semiconductor parameter analyzer 4156C).

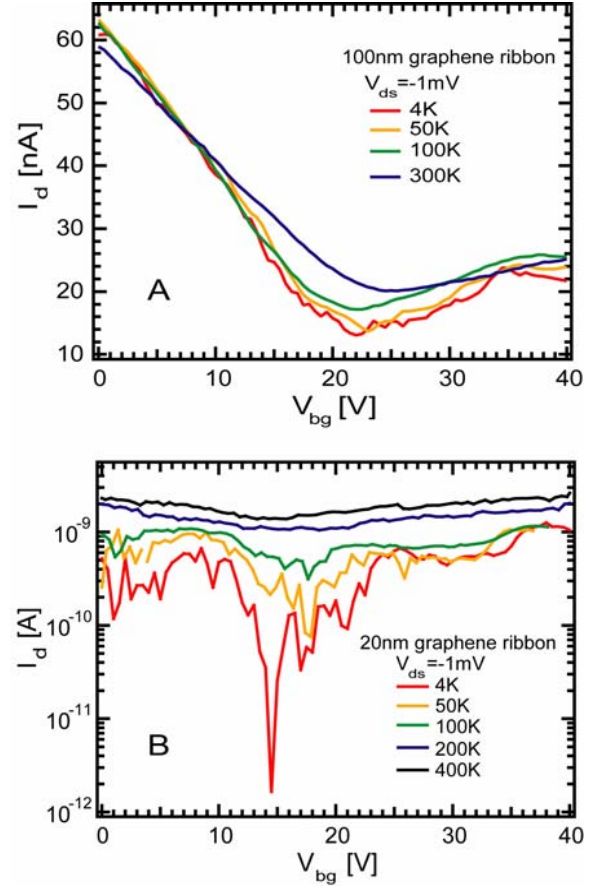

Fig. 3: Temperature dependence measurements: (A) *100 nm* GNR (B) *20 nm* GNR. $V_{ds}=-1mV$ was used in all measurements. The minimum current of the *100 nm* GNR device decreases less than factor of 2 from *300 K* to *4 K*, while the same drops more than 1.5 orders of magnitude for a *20 nm* GNR.

In order to quantitatively analyze how large the gap is, we measured the *20 nm* GNR at various temperatures and created an Arrhenius plot of the minimum current, as shown in Fig. 4. In an FET with an intrinsic semiconductor channel and a mid-gap line-up with the source/ drain metal, the lowest current point of the FET is reached when both, the conduction and valence band are flat. At this stage, the current is dominated by thermally activated carriers and follows an exponential dependence on temperature: $I_{off} \propto exp(-E_g/2k_BT)$. At high temperatures, a clear linear relation between $logI_{off}$ and $1/T$ is observed, and a gap of *28 meV* is extracted for the *20 nm* GNR. This value is in good agreement with the predictions of a recent density functional theory study of GNRs [19]. The *4 K* data point falls out of the linear dependence. It is understood that at such low temperatures, carrier transport is limited by tunneling rather than thermal injection.

Strong current fluctuations were observed at low temperatures for narrow GNRs, as shown in Fig. 3 (B). Those current variations are possibly due to universal conductance fluctuation, which may indicate the presence of elastic scattering centers. Due to instrumental limitation, the scanning step of the gate voltage and the applied bias voltage are both too large to accurately extract the amplitude of the fluctuations. Further measurements are needed to quantitatively analyze and conclude about the scattering mechanism responsible for the conductance fluctuation.

As discussed earlier, it is very likely that mixed edge states exist in our graphene ribbons. This mixture will cause different boundary confinements in different graphene segments – some segments may become semiconducting while others stay metallic. Recent first-principles calculations indicate that the gap is also influenced by the edge states [20]. This complexity makes the analysis of the real device geometry, such as the channel length, much more difficult. However, these different segments are most likely connected to each other in series, therefore, the transport characteristics shown in Fig. 3 (B) are dominated by the semiconducting components.

current to more negative gate voltages and the appearance of the electron branch instead of the hole branch in the same gate voltage window. Due to the aforementioned impact by trapped charges in the $SiO_2$ substrate, the shift of the Fermi level cannot be quantitatively determined from the threshold voltage difference. However, a roughly *50 V* shift observed in the characteristics is much larger than the substrate effect and indeed indicates a successful doping. In K-doped CNT devices, a combination of bulk doping and contact effects has been considered to be responsible for the observed transport properties. The increase of Schottky barrier height to the valence band (or equivalently, a decrease to the conduction band) results in a suppression of the hole branch [21]. However, this is not the case for GNR devices. Even for our smallest ribbon with a finite semiconducting gap, the gap is so small that barriers at the contact interface play a minor role at room temperature. Therefore, if the gate voltage can be expanded without gate leakage, both hole and electron branches should be observed with similar current levels and only the threshold voltage shift is the indication of the doping in GNR devices.

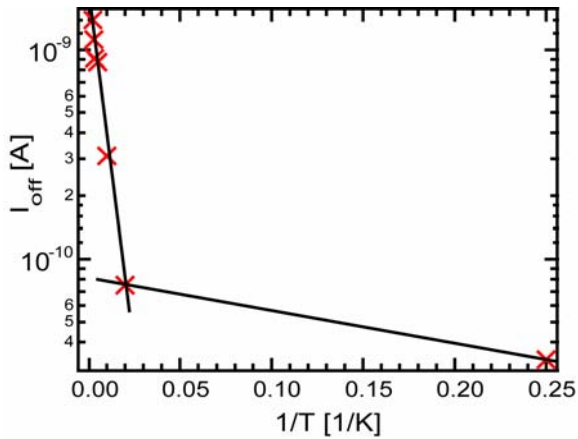

Fig. 4: Off state current as a function of inverse temperature for a *20 nm* GNR. $V_{ds}=-1mV$ for all data points.

Next we compare the characteristics of our original GNR devices with the same devices after potassium (K) doping and annealing. The details of the doping process can be found in ref. [21]. As shown in Fig. 5, within the same gate voltage range, before K doping, a *30 nm* GNR device shows a strong hole branch, while the same device shows a clear electron branch after doping. Relying on charge transfer, potassium acts as an electron donor and shifts the Fermi level of the GNR into the conduction band. This directly results in the observed shift of the minimum

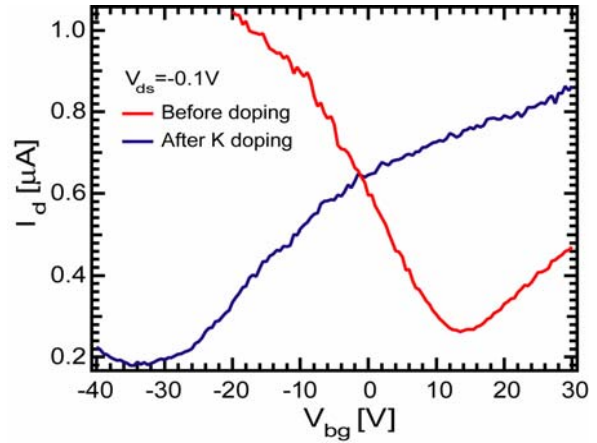

Fig. 5: Characteristics of a *30 nm* GNR device before and after potassium doping. Measurements were taken at room temperature.

In evaluating the merits of GNR devices for potential technological applications, the electrical noise behavior, which determines the signal-to-noise ratio, is one of the essential factors that should be investigated and considered. It is well known that current fluctuations (current noise), increase as the device dimension shrinks. In nano-scaled devices, such as carbon nanotubes, the current noise is usually dominated by the so-called *1/f* noise that exhibits a power spectrum inversely proportional to the frequency *f* [22,23]. Due to the structure similarities between carbon nanotubes and GNRs, it is also interesting to compare their noise characteristics.

We have measured the electrical current noise in GNR devices, and found that the current fluctuation is indeed dominated by *1/f* noise at low frequencies (see Fig. 7). As shown in Fig. 6, the current noise power density $S_I$ of GNRs exhibits an $I^2$-dependence, and can be described by: $S_I = A(I^2/f)$, where *A* refers to the *1/f* noise amplitude. This $I^2$ dependence indicates that the current fluctuation is a manifestation of the resistance fluctuation. In carbon nanotubes, the noise amplitude *A* was found to be related to number of transport carriers *N* in the system with *A* being inversely proportional to *N* [23,24]. In order to evaluate the impact of the edge states to the *1/f* noise behavior in GNR devices, we have measured the noise power spectra of two GNR devices with very different widths (*20 nm* and *200 nm*), as shown in Fig. 7. At $V_g=0$, the resistance *R* of the *20 nm* device is *670 k$\Omega$* and that of the *200 nm* device is *30 k$\Omega$*. The noise amplitudes *A* are $3\times10^{-5}$ and $3\times10^{-6}$ for the *20 nm* and *200 nm* devices, respectively, showing good agreement with the *A* ~ *1/N* behavior. Therefore, while the GNR devices exhibit a width-dependent resistivity due to contributions from edge states (see Fig. 2), the fact that the noise amplitude *A* follows the *1/N* relation suggests that the *1/f* noise is not significantly affected by the presence of the edge states.

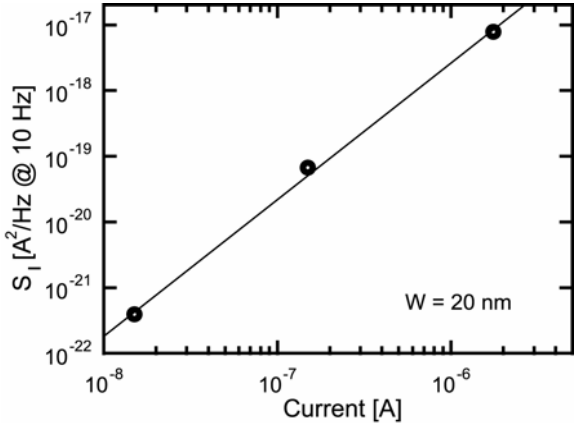

Fig. 6: Current noise power $S_I$ at *10 Hz* as a function of dc bias current *I* for a *20 nm* wide graphene device. The measurements were performed at zero gate voltage, $V_g = 0V$, with drain voltages $V_{ds} = 10\ mV$, *100 mV* and *1 V*. The current noise $S_I$ fits an $I^2$ behavior.

We have also compared the amplitude of the *1/f* noise of GNR and CNT devices with the same channel length. In carbon nanotubes, the noise amplitude *A* can be described by the empirical relation of $A/R = 9\times10^{-11}$ ($\Omega^{-1}$) for as-prepared CNT devices [25], which is very close to the *A/R* values (~$5-10\times10^{-11}\ \Omega^{-1}$) of our GNRs. These results further suggest that, similar to the case of CNT devices [12,23], the *1/f* noise in GNR devices is mainly associated with extrinsic fluctuation mechanisms, such as fluctuations in the occupancy of charged traps in the gate oxide (substrate), instead of carrier scattering processes such as intrinsic impurity-induced scattering, or phonon scattering.

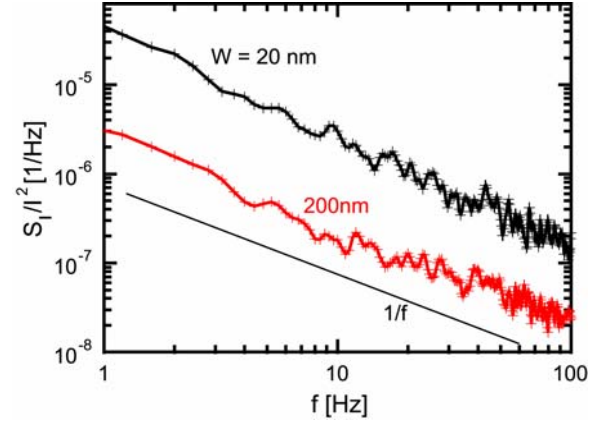

Fig. 7: Normalized current noise power spectrum $S_I/I^2$ for two graphene devices with different widths, *20 nm* and *200 nm*. Both devices have the same channel length of *1 μm*. The noise spectrum shows the characteristic *1/f* dependence. For the *200 nm* device, the noise amplitude is about 10 times lower than the *20 nm* device, in agreement with *1/N* behavior, indicating that the scattering associated with edge states does not contribute to the generation of *1/f* noise.

## 4. Summary

In conclusion, we have shown that GNRs as narrow as *20 nm* can be fabricated by e-beam lithography and etching techniques and be incorporated as channels of field effect transistors. We have found that both boundary scattering and trapped charges in the substrate strongly affect the transport properties and minimum conductivity of the GNRs. A confinement-induced gap of the order of *30 meV* was inferred in the narrowest *20 nm* ribbon. The dominant electrical noise at low frequencies was found to be *1/f* noise arising from fluctuations in the occupancy of charged traps in the substrate.

### Acknowledgement

The authors would like to thank Dr. Joerg Appenzeller for stimulating discussions, Bruce Ek for expert technical assistance, and Dr. Xu Du from Rutgers University for the supply of HOPG and helpful discussions.

### References

* Corresponding author. E-mail: zchen@us.ibm.com